# Cumulus parameterization controls simulated rainfall more than microphysics, but not at heavy thresholds: a 3 × 2 factorial WRF study over the Western Himalaya

Satyam Kumar[1]*, Akshay Sunil[2], Devesh Sheoran [3]

[1] School of Earth, Ocean and Climate Sciences, Indian Institute of Technology Bhubaneswar, Argul, Khordha, Odisha 752050, India.

[2] Centre for Climate Studies, Indian Institute of Technology Bombay, Mumbai, Maharashtra 400076, India.

[3] School of Earth, Ocean and Climate Sciences, Indian Institute of Technology Bhubaneswar, Argul, Khordha, Odisha 752050, India.

Email: Satyam Kumar: satyam72kr@gmail.com

Akshay Sunil: akshay.sunil@experiqs.tech

Devesh Sheoran: deveshsheoran000@gmail.com

* Corresponding author: Satyam Kumar, satyam72kr@gmail.com

## Abstract

Rainfall forecasts over the Western Himalaya remain poor, and the relative contributions of the cumulus and microphysics parameterizations to that error are not well constrained. Nine WRF simulations of the 6–13 July 2023 Western Himalayan heavy-rainfall event were run at 9 km grid spacing over a common domain, differing in exactly two namelist options; six of them form a complete 3 × 2 factorial in microphysics (WSM6, Thompson, WDM6) and cumulus (Kain–Fritsch, Tiedtke). Verification used IMD 0.25° gridded rainfall over 447 land points for 7–12 July 2023. Switching cumulus scheme changes the domain-mean bias by 10.5 mm day$^{-1}$ against 1.4 mm day$^{-1}$ for microphysics, a factor of 7.8; the corresponding ratios are 4.2 for daily spatial correlation and 4.8 for detection of rainfall above 35 mm day$^{-1}$. Kain–Fritsch is systematically wet (+9.3 to +11.0 mm day$^{-1}$) but better located, whereas Tiedtke attains near-zero domain-mean bias only by cancelling a northern excess against a central deficit. The cumulus advantage is score-dependent at heavy thresholds: at 35 mm day$^{-1}$ and above the Fractions Skill Score favours Tiedtke at neighbourhood widths of three grid points or more, while detection of heavy rainfall continues to favour Kain–Fritsch. Matching the model accumulation window to the IMD reporting day alters verified skill by as much as the physics choice itself.

## 1. Introduction

Quantitative precipitation forecasting over the Western Himalaya is among the harder problems in regional numerical weather prediction. The rainfall is set by moist monsoonal and extratropical flow meeting terrain that rises from the Indo-Gangetic plain to above 6000 m within less than 200 km. Satellite climatologies resolve narrow bands of heavy precipitation tied to breaks in slope rather than to elevation itself (Bookhagen and Burbank 2006), and the convective systems that deliver the rain differ in structure from those over the adjacent plains, being shallower, more frequent and organised by the terrain and its diurnal cycle (Medina et al. 2010; Romatschke and Houze 2011). Little of this structure is resolved at operational grid spacings, the rainfall is organised on scales close to the model grid, and the observational network that would constrain either the initial state or the verification is thin at elevation, with gauge density falling away where the rainfall gradients are steepest (Dimri et al. 2004; Patil and Pradeep Kumar 2016).

The practical stakes are considerable. The frequency of extreme daily rainfall over India rose over the second half of the twentieth century (Goswami et al. 2006), and widespread extreme events over central India have increased roughly threefold since 1950 (Roxy et al. 2017). Over the Himalaya the consequences are amplified, because short-duration, high-intensity rainfall on steep and weathered slopes converts into debris flow and flash flood with little warning time (Dimri et al. 2017). Several individual events have been costly enough to warrant case studies in their own right, among them the 2010 Leh cloudburst (Rasmussen and Houze 2012; Thayyen et al. 2013) and the 2013 Uttarakhand disaster (Chevuturi and Dimri 2016).

Regional models are the only practical tool for anticipating such events, and their behaviour over this terrain has been documented in some detail. Year-long WRF integrations over the Himalaya reproduce the large-scale seasonal cycle but misplace the elevation of maximum precipitation and the phase of its diurnal cycle (Norris et al. 2017), and refining the grid into the convection-permitting range improves the diurnal cycle and the representation of heavy rainfall while leaving substantial biases in place (Karki et al. 2017). Physics choice remains a first-order control on the result at every resolution so far examined.

A large body of work has addressed this by tuning physics parameterizations, and the general finding — that simulated rainfall is sensitive to the choice of cumulus and microphysics schemes — is well established for Indian domains (Patil and Pradeep Kumar 2016; Dixit et al. 2023; Sharma et al. 2024; Biswasharma et al. 2024). What is much less clear is the *relative* magnitude of the two sensitivities. Most published sensitivity studies vary several physics options simultaneously, or vary one group while holding a single member of the other group fixed. Neither design allows the effect of cumulus scheme to be separated cleanly from the effect of microphysics scheme, because the two are not orthogonal in the experiment matrix. A practitioner who must decide where to spend limited configuration effort is therefore left without a quantitative answer to a simple question: which choice matters more?

The closest existing design to the one used here is that of Dixit et al. (2023), who crossed three microphysics schemes with two cumulus schemes over a Himalayan headwater catchment. That study was a year-long precipitation climatology aimed at identifying a seasonally suitable configuration, and it did not quantify the two main effects against each other. The present study takes the complementary approach: a single extreme event, a strictly orthogonal factorial, and a scale-dependent verification framework, so that the relative magnitude of the two effects can be stated as a number.

The event chosen for that comparison is the extreme rainfall of early July 2023, which produced exceptional totals over Himachal Pradesh and Uttarakhand. The India Meteorological Department recorded large excess rainfall over Himachal Pradesh from 8 to 12 July, with daily station totals of 228.5 mm at Una Rampur and 224.0 mm at Bilaspur on 9 July and 250.0 mm at Nahan on 11 July, against monthly totals of 400–700 mm across the state (IMD 2023; Sana et al. 2024). The rainfall triggered flash floods and widespread slope failure: Sana et al. (2024) mapped approximately 6500 landslides over an area of about 3100 $km^2$ in the Mandi and Kullu districts during the 2023 monsoon season, and the state's post-disaster needs assessment documents the resulting loss of life and the damage to housing, roads and agriculture (HPSDMA 2023).

The synoptic setting was an interaction between a slow-moving, westward-propagating active western disturbance and the monsoonal trough (Sana et al. 2024). Western disturbances are the dominant synoptic systems of the winter season over northwest India, but they occur throughout the year, and their warm-season occurrences are the ones most often implicated in Himalayan extreme rainfall (Dimri et al. 2015). Such transient merging between extratropical and monsoonal systems is a recognised mechanism for heavy precipitation over South Asia (Hunt et al. 2021). The interaction draws a deep, moist monsoonal layer against the terrain while an upper-level trough supplies the ascent to release it, a configuration diagnosed for Himalayan extreme rainfall episodes generally (Vellore et al. 2016) and for the 2010 Pakistan floods in particular (Houze et al. 2011). Observed domain-mean rainfall over the analysis area averaged 23.4 mm day$^{-1}$ over 7–12 July, with a single-point event total of 627 mm.

The event is well suited to a physics-sensitivity study for three reasons. It is large enough that the rainfall signal is not dominated by observational noise; it spans several days, so verification is not hostage to a single day's timing error; and it produced heavy-tail rainfall, allowing the schemes to be discriminated at thresholds that matter operationally.

The grid spacing at which the event is simulated places the model in a particular dynamical regime, and that regime is what makes the comparison interesting. At 9 km the model sits in the convective grey zone, the range of resolutions at which a process is partially resolved and the assumptions behind its parameterization are no longer strictly valid (Wyngaard 2004; Gerard 2007). Deep moist convection is not adequately resolved until grid spacings well below 1 km (Bryan et al. 2003), so at 9 km convection is neither fully resolved nor cleanly separable from the parameterized component. Cumulus parameterization is therefore still applied, but its assumptions are strained. This makes the grey zone precisely the regime in which the relative weight of cumulus against microphysics is both least obvious a priori and most consequential in practice, since operational regional configurations over India commonly run at comparable resolutions. How models behave as they cross into and through this range has been reviewed for convection-permitting simulation generally (Prein et al. 2015), and the particular question of how cumulus and microphysics parameterizations perform across the grey zone has been examined for other regions (Jeworrek et al. 2019), but not, as far as we are aware, in an orthogonal factorial design over Himalayan terrain.

A second consideration concerns how such a comparison is assessed rather than how it is run. Comparing a rainfall forecast with a gridded analysis point by point penalises a displaced feature twice: once as a miss where the rain was observed, and again as a false alarm where it was simulated. The penalty grows as resolution is refined, so a model that produces more realistic small-scale structure can score worse than a smoother one that produces none. Neighbourhood, or fuzzy, verification methods were developed to separate this displacement error from an error of amount or frequency, by relaxing the requirement for an exact match over a range of spatial scales (Ebert 2008; Gilleland et al. 2009). The Fractions Skill Score (Roberts and Lean 2008) is the most widely used member of that family and reports, as a function of neighbourhood size, the scale at which a forecast first becomes useful. It is adopted here because the question under study is itself scale-dependent: two parameterization groups may differ little in area-average rainfall while differing substantially in the scale at which that rainfall is correctly placed.

Against this background, the present study asks three questions. First, in a strictly controlled experiment, how much larger is the effect of the cumulus scheme than the effect of the microphysics scheme on simulated rainfall? Second, does that answer depend on the rainfall threshold and on the spatial scale at which skill is assessed? Third, how much of the apparent model error is attributable to verification methodology rather than to model physics? The third question is not usually posed alongside the first two, but it proved unavoidable: two routine processing choices changed the verified skill by more than the difference between some of the schemes being compared.

## 2. Data and methods

This section sets out the model configuration and the verification framework. Sections 2.1 to 2.3 describe what was run and where it is assessed: the common model configuration and the nine physics combinations, the observational reference, and the analysis domain. Sections 2.4 and 2.5 deal with two processing steps that are normally treated as routine and reported in a single sentence — the alignment of the model accumulation window with the IMD reporting day, and the interpolation of model output onto the observation grid. They are documented here in more detail than is customary because, as Sect. 3 shows, each changed the verified skill by an amount comparable to the physics difference under study, and a reader attempting to reproduce or extend this work would obtain different numbers without them. Sections 2.6 to 2.8 define the verification metrics, the factorial analysis and the software used.

### 2.1 Model configuration

All simulations used the Advanced Research WRF (ARW) core, version 4.5 (Skamarock et al. 2019). A single domain of 81 × 81 mass points at 9 km grid spacing was centred on 31.0°N, 78.5°E on a Lambert conformal projection with standard parallels at 30°N and 34°N, giving a domain span of 720 km. Forty-five terrain-following vertical levels were used with a model top at 50 hPa.

The domain, its terrain and the verification region are shown in Fig. 1.

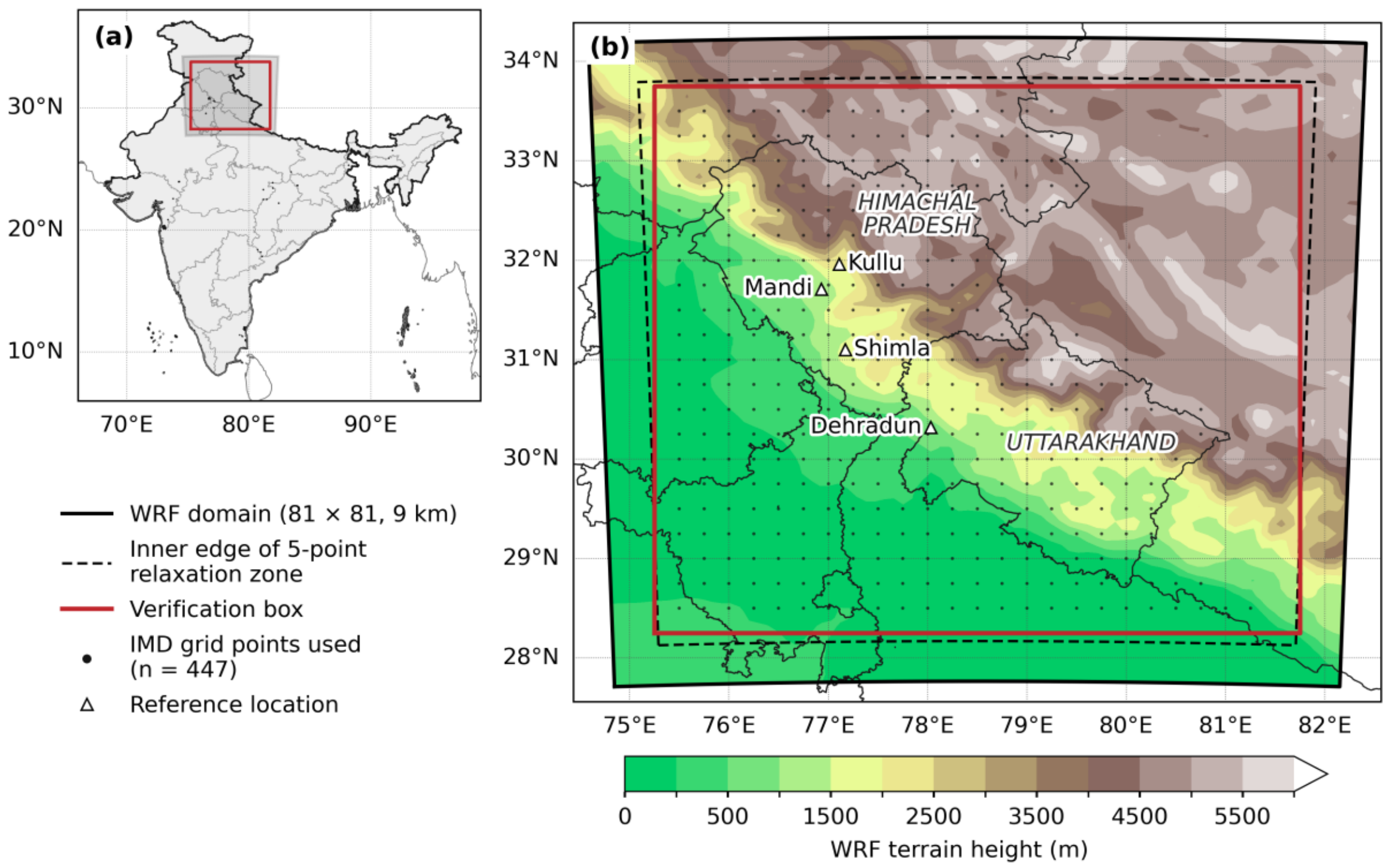

**Fig. 1** Study region. **a** Location of the WRF domain (grey) and the verification box (red) over India. **b** WRF terrain height over the 81 × 81, 9 km model domain (solid black line), with the inner edge of the five-point lateral boundary relaxation zone (dashed), the verification box at 28.25–33.75°N, 75.25–81.75°E (red) and the 447 IMD 0.25° grid points used for verification (dots). Triangles mark reference locations

Initial and lateral boundary conditions were taken from NCEP FNL 1° operational global analyses at 6-hourly intervals (NCEP 2000), with 34 metgrid levels. The lateral boundary relaxation zone width (spec_bdy_width) was five grid points. Integration ran from 0000 UTC 5 July to 1800 UTC 13 July 2023 with a 54 s time step, writing hourly output.

Nine experiments were performed (Table 1).

**Table 1** Experiment matrix. Only mp_physics and cu_physics differ across the nine namelists.

| Exp | mp_physics | Microphysics | cu_physics | Cumulus | In factorial |
|---|---|---|---|---|---|
| 1 | 6 | WSM6 | 1 | Kain–Fritsch | yes |
| 2 | 8 | Thompson | 6 | Tiedtke | yes |
| 3 | 8 | Thompson | 3 | Grell–Freitas | no |
| 4 | 16 | WDM6 | 6 | Tiedtke | yes |
| 5 | 6 | WSM6 | 6 | Tiedtke | yes |
| 6 | 8 | Thompson | 1 | Kain–Fritsch | yes |
| 7 | 16 | WDM6 | 1 | Kain–Fritsch | yes |
| 8 | 6 | WSM6 | 16 | New Tiedtke | no |
| 9 | 3 | WSM3 | 5 | Grell 3D ensemble | no |

The nine namelists were compared option by option and differ in exactly two entries, mp_physics and cu_physics. All 75 remaining options are identical across every run, including RRTM longwave radiation (Mlawer et al. 1997), Dudhia shortwave radiation (Dudhia 1989), the YSU planetary boundary layer scheme (Hong et al. 2006), the MM5 similarity surface layer, the Noah land surface model (Chen and Dudhia 2001), and the gravity wave drag option. No nesting was used.

Experiments 1, 5, 6, 2, 7 and 4 constitute a complete 3 × 2 factorial in microphysics (WSM6, Thompson, WDM6) crossed with cumulus (Kain–Fritsch, Tiedtke). Experiments 3, 8 and 9 extend the cumulus sampling to Grell–Freitas, New Tiedtke and the Grell 3D ensemble scheme but do not participate in the factorial analysis.

Microphysics references are Hong et al. (2004) for WSM3, Hong and Lim (2006) for WSM6, Thompson et al. (2008) for Thompson, and Lim and Hong (2010) for WDM6. Cumulus references follow the attribution given for each namelist option in the WRF Users Guide: Kain (2004) for Kain–Fritsch (`cu_physics` = 1), Grell and Freitas (2014) for Grell–Freitas (3), Grell (1993) with Grell and Dévényi (2002) for the Grell 3D ensemble scheme (5), Tiedtke (1989) with Zhang et al. (2011) for the Tiedtke scheme as implemented in WRF (6), and Zhang and Wang (2017) for New Tiedtke (16).

### 2.2 Observations

Verification used the India Meteorological Department 0.25° × 0.25° gridded daily rainfall product (Pai et al. 2014), which is interpolated from the national rain-gauge network. The GPM IMERG Final Run daily product, version 07B, at 0.1° (Huffman et al. 2023), conservatively remapped to the IMD grid, together with ERA5 (Hersbach et al. 2020) and GFS analyses, was used qualitatively for cross-comparison of the event's daily evolution but not for quantitative scoring. IMERG accumulates on the UTC calendar day, which is why it peaks one day before IMD and the model fields in Fig. 5.

### 2.3 Analysis domain

The analysis box was set to 28.25–33.75°N, 75.25–81.75°E, giving a 23 × 27 grid.

This is deliberately smaller than the full model domain. With a five-point relaxation zone, the interior free of lateral boundary relaxation extends 315 km from the domain centre. A box spanning 28–34°N, 75–82°E reaches 344 km from the centre, placing approximately 11% of its area inside the relaxation rim, where the solution is nudged towards the coarse driving analysis. Trimming to the values above moves the verification region into the free interior at a cost of one grid row and column on each side, with one small exception: at the southern end of its eastern and western edges the box still reaches up to 3.7 km, that is 0.4 of a model grid point, into the innermost row of the relaxation zone, where the nudging is weakest (Fig. 1b).

Of the 621 cells in the trimmed box, 447 have valid IMD data on every verification day; the remainder lie outside the gauge network's coverage, largely on the Tibetan side of the range. All statistics use this common mask, applied identically to observations and to every experiment. The verification domain therefore covers about 72% of the trimmed box and about 55% of the model domain area.

### 2.4 Temporal alignment

WRF accumulates rainfall on the UTC calendar day. The IMD gridded product does not: each daily value represents the 24 hours ending at 0830 IST, that is 0300 UTC (Pai et al. 2014), the window IMD also uses in its daily rainfall bulletins (IMD 2023). Critically, the accumulation from 0830 IST of one day to 0830 IST of the next is reported as the daily rainfall of the *later* day (Ponukumati et al. 2023). The label therefore corresponds to the date on which the gauge was read, not the date on which most of the rain fell, and comparing the two products by date misaligns them by close to one day.

The misalignment was first detected visually. Domain-mean rainfall in IMD peaked on 9 July at 50 mm day$^{-1}$, whereas IMERG, ERA5, GFS and all nine WRF experiments peaked on 8 July. That a satellite product, a reanalysis, a global analysis and nine independently configured simulations all agreed with one another and disagreed with a single gauge-based product pointed to a labelling convention rather than to a physical error. The 9 July IMD peak is itself consistent with the labelling: Sana et al. (2024) give 9 July as the day of maximum recorded station rainfall for this event.

The required shift can be derived rather than fitted. The hourly model output is de-accumulated with hour-ending timestamps, so the value at 0100 UTC is the rainfall between 0000 and 0100 UTC, and a plain daily sum bins the window 2300–2300 UTC. Applying a forward time shift of $N$ hours before the daily sum makes the bin labelled $X$ cover $(X00{:}00 - N - 1\text{ h}, X23{:}00 - N]$. Setting this equal to the IMD window, 0300 UTC of $X-1$ to 0300 UTC of $X$, gives $N = 20$ h.

This prediction was tested by rebuilding the daily model accumulation with a range of shifts and rescoring experiment 1 against an identical set of IMD days (Table 2). RMSE and daily spatial correlation both attain their optimum at exactly the shift predicted above, and degrade on either side of it, at +22 h and +24 h as well as at −24 h. The response is therefore a physical optimum at the documented reporting window rather than an artefact of shifting as such, and the derivation preceded the test. The probability of detection at 35 mm day$^{-1}$ prefers +22 h marginally, but it is a threshold-crossing count and is noisier than the two full-distribution scores, which agree. A shift of +20 h was adopted for all subsequent analysis.

**Table 2** Sensitivity of experiment 1 verification scores to the model accumulation window. Bias and RMSE in mm day$^{-1}$. Verification period 7–12 July 2023 in every row, so all rows use identical observations and the same post-spin-up model data.

| Shift | Window (UTC) | Bias | RMSE | Daily r | POD 35 |
|---|---|---|---|---|---|
| −24 h | 2300–2300, one day early | 8.37 | 54.84 | 0.180 | 0.362 |
| none | 2300–2300 | 10.20 | 43.64 | 0.375 | 0.576 |
| +20 h | 0300–0300 (IMD window) | 9.57 | 36.34 | 0.437 | 0.669 |
| +22 h | 0100–0100 | 9.48 | 36.47 | 0.427 | 0.680 |
| +24 h | 2300–2300, one day late | 9.38 | 37.05 | 0.411 | 0.684 |

Bias is only weakly sensitive to the shift, ranging from 8.4 to 10.2 mm day$^{-1}$ across a 44-hour range, whereas RMSE and correlation are strongly sensitive. This is expected, because bias is a multi-day domain average and is little affected by moving rainfall between days, whereas point-by-point agreement is not. It also means that the principal conclusion of this study, which concerns bias, is robust to this choice.

Because the forward shift draws part of the 6 July accumulation window from the spin-up period, the verification window was set to 7–12 July 2023, six days. 5 July is spin-up and 13 July is excluded because the run terminates at 1800 UTC, before the IMD accumulation window for that date closes.

### 2.5 Spatial regridding

Model rainfall was regridded from the native 9 km grid to the IMD 0.25° grid using bilinear interpolation.

An earlier version of this analysis used bicubic interpolation. Bicubic weights are not sign-constrained and overshoot across sharp gradients; in this domain that produced accumulations as low as −3.89 mm day$^{-1}$, which is physically impossible and implies compensating positive overshoot elsewhere. Since the sharp gradients here coincide with the orographic rainfall maxima, the affected regions are precisely those carrying the signal. Bilinear interpolation, whose weights are non-negative and sum to unity, cannot produce values outside the range of the surrounding source points; after switching, the minimum across all nine experiments was of order $10^{-10}$ mm day$^{-1}$, that is floating-point noise.

Bilinear interpolation is not conservative and smooths the field, so the regridded model maxima are slightly reduced relative to the native grid. The effect is modest here because the target grid at 0.25° is coarser than the 9 km source grid by only a factor of about three, and it acts in the same direction for all nine experiments, so it does not affect the between-scheme comparison that is the subject of this study. Conservative remapping would be preferable in principle, because it would preserve the area-integrated rainfall exactly, but the WRF output grid is stored as a curvilinear grid without cell-corner coordinates, and the software build available did not support generating them.

### 2.6 Verification metrics

Standard scores were computed over the 2682 point-days: mean bias, RMSE and Pearson correlation. Two correlations are reported. The daily spatial correlation is computed across the 447 valid points separately for each day and then averaged, which is what is normally meant by spatial correlation in this context; a pooled correlation over all point-days is also computed but mixes spatial and temporal variance and is not used for the main conclusions.

Categorical scores were computed at 1, 10, 20, 35 and 50 mm day$^{-1}$.

The Fractions Skill Score (Roberts and Lean 2008) was computed at the same thresholds for square neighbourhoods of 1, 3, 5, 7 and 9 grid points. On the 0.25° verification grid one grid point corresponds to 27.8 km in latitude and about 23.8 km in longitude at 31°N, so a neighbourhood is not exactly square in physical units. Widths are quoted in grid points throughout; where a physical width is given it uses the latitudinal spacing of 27.8 km per point, so that a three-point neighbourhood is quoted as approximately 83 km. Neighbourhood fractions are normalised by the number of *valid* points in each window rather than by the window area, so that masked cells are excluded from both numerator and denominator; this matters near the northern edge of the box, where up to about 30% of a large window can be invalid. The useful-skill target is taken as $0.5 + f_0/2$, where $f_0$ is the observed base rate for that threshold.

### 2.7 Factorial analysis

The six factorial experiments form a complete 3 × 2 design: every microphysics scheme is run with every cumulus scheme, and no other namelist option varies. This orthogonality is what makes the two effects separable. Where several options change together, or where one group is varied while a single member of the other is held fixed, the effect attributed to one group is confounded with the particular choice made for the other, and the resulting sensitivity is conditional on that choice in a way that cannot be quantified from the experiment set itself.

For each score, the marginal mean of a level is the average over the experiments sharing that level. The main effect of a factor is the range of its marginal means. For cumulus, which has two levels, this is the difference between the mean over the three Kain–Fritsch runs and the mean over the three Tiedtke runs; for microphysics, which has three, it is the range of the WSM6, Thompson and WDM6 marginal means. Their ratio quantifies how much more the cumulus choice matters than the microphysics choice, with all other configuration options held identical by construction.

The design also allows the interaction between the two factors to be examined, that is, whether the effect of changing microphysics depends on which cumulus scheme is in use. The interaction residual in each cell is the difference between the simulated value and the value predicted by an additive model built from the grand mean and the two sets of marginal means. With a single realisation per cell there is no replicate error against which the interaction could be tested for significance, but its magnitude can be compared with the two main effects, which is sufficient to establish whether the main effects can be read as separate quantities. That comparison is reported in Sect. 3.2.

### 2.8 Software and tools

Hourly model output was de-accumulated, time-shifted, summed to daily totals and interpolated to the observation grid with Climate Data Operators (CDO 2.x), using the `shifttime`, `daysum` and `remapbil` operators for the model fields and `remapcon` for the IMERG product, which is on a regular grid and therefore admits conservative remapping. Verification statistics, the factorial main effects, the Fractions Skill Score and all figures were produced with Python 3 using xarray and netCDF4 for input, NumPy and pandas for the statistics, the uniform filter of `scipy.ndimage` for the neighbourhood fractions, and Matplotlib, Cartopy and GeoPandas for the figures. Administrative boundaries in the map figures are drawn from a single shapefile applied identically to every panel.

## 3. Results

The central result is that the nine experiments separate by cumulus scheme and not by microphysics scheme, and that this separation is large, systematic and — with one qualification at heavy rainfall thresholds — consistent across the scores examined. The remainder of this section establishes that result and then tests it from several directions.

Section 3.1 presents the domain-mean statistics, where the separation is first visible. Section 3.2 quantifies it formally as a ratio of factorial main effects and examines whether the two effects are additive. Section 3.3 asks whether the separation survives when skill is assessed as a function of spatial scale and rainfall threshold, where it is found to weaken and then reverse. Sections 3.4 and 3.5 turn from the factorial to the individual configurations, including the three that lie outside it. Section 3.6 examines the spatial fields, which are needed to explain why two configurations with almost identical domain-mean bias are not equally skilful, and which also provide an independent check on the temporal alignment of Sect. 2.4.

### 3.1 Domain-mean statistics

Verification statistics for all nine experiments are given in Table 3. Observed domain-mean rainfall over the verification period is 23.40 mm day$^{-1}$.

The experiments separate cleanly by cumulus scheme rather than by microphysics scheme. All three Kain–Fritsch runs are wet, by 9.32 to 10.99 mm day$^{-1}$; all three Tiedtke runs are close to neutral, between −1.13 and −0.10 mm day$^{-1}$; the New Tiedtke run behaves like the Tiedtke runs at −1.25 mm day$^{-1}$; and the two Grell-family runs sit apart from both, at +1.94 mm day$^{-1}$ for Grell–Freitas and −5.01 mm day$^{-1}$ for the Grell 3D ensemble scheme. Within each cumulus group the spread introduced by changing microphysics is an order of magnitude smaller than the difference between groups: 1.67 mm day$^{-1}$ across the three Kain–Fritsch runs and 1.03 mm day$^{-1}$ across the three Tiedtke runs, against 10.52 mm day$^{-1}$ between the groups.

The same grouping appears in the other scores. Daily spatial correlation spans 0.294 to 0.449 across the nine runs, with the three Kain–Fritsch runs occupying a narrow band at the top of that range, 0.437 to 0.449, and the four Tiedtke-family runs the lower part, 0.294 to 0.357. Detection of rainfall above 35 mm day$^{-1}$ separates further still: 0.669 to 0.699 for Kain–Fritsch against 0.458 to 0.537 for the Tiedtke family. In both cases the three microphysics schemes within a cumulus group fall closer to one another than any Kain–Fritsch run falls to any Tiedtke run.

RMSE behaves differently and is treated separately. Within the factorial the ordering still holds, with all three Kain–Fritsch runs below all three Tiedtke runs, but the two lowest values of the nine belong to non-factorial runs: experiment 9 at 29.83 mm day$^{-1}$ and experiment 3 at 34.26 mm day$^{-1}$. Experiment 9 is also the driest run in the set, which is the first indication — developed in Sect. 3.5 — that a low RMSE on a heavy-tailed field can be obtained by producing too little rain rather than by placing it better.

An ordering this consistent, produced by experiments that differ in a single namelist entry, is what motivates the formal factorial treatment that follows.

**Table 3** Verification statistics for all nine experiments, 7–12 July 2023, 447 land points, 2682 point-days. Model mean, bias and RMSE in mm day$^{-1}$. Observed mean 23.40 mm day$^{-1}$.

| Exp | Microphysics | Cumulus | Model | Bias | RMSE | Daily r | POD 1 | POD 10 | POD 35 |
|---|---|---|---|---|---|---|---|---|---|
| 1 | WSM6 | Kain–Fritsch | 32.97 | +9.57 | 36.34 | 0.437 | 0.957 | 0.818 | 0.669 |
| 2 | Thompson | Tiedtke | 23.29 | −0.10 | 38.25 | 0.294 | 0.788 | 0.587 | 0.508 |
| 3 | Thompson | Grell–Freitas | 25.34 | +1.94 | 34.26 | 0.416 | 0.912 | 0.697 | 0.559 |
| 4 | WDM6 | Tiedtke | 22.27 | −1.13 | 36.07 | 0.357 | 0.771 | 0.570 | 0.537 |
| 5 | WSM6 | Tiedtke | 22.95 | −0.45 | 37.46 | 0.352 | 0.763 | 0.578 | 0.533 |
| 6 | Thompson | Kain–Fritsch | 34.38 | +10.99 | 35.50 | 0.449 | 0.962 | 0.831 | 0.699 |
| 7 | WDM6 | Kain–Fritsch | 32.71 | +9.32 | 33.91 | 0.437 | 0.949 | 0.819 | 0.674 |
| 8 | WSM6 | New Tiedtke | 22.15 | −1.25 | 38.90 | 0.308 | 0.810 | 0.577 | 0.458 |
| 9 | WSM3 | Grell 3D ens. | 18.39 | −5.01 | 29.83 | 0.401 | 0.927 | 0.664 | 0.412 |

### 3.2 The factorial result

Domain-mean bias for the 3 × 2 factorial is given in Table 4 and shown in Fig. 2. Main effects are defined here in the standard factorial sense: for each factor, the six experiments are averaged over the levels of the other factor to give marginal means, and the effect is the range of those marginal means. For cumulus, which has two levels, this is the difference between the mean of the three Kain–Fritsch runs and the mean of the three Tiedtke runs, 10.52 mm day⁻¹. For microphysics, which has three levels, it is the range of the WSM6, Thompson and WDM6 marginal means, 1.35 mm day⁻¹. The cumulus scheme therefore controls the domain-mean bias 7.8 times more strongly than the microphysics scheme, with every other configuration option held identical.

The comparison is conservative in one respect. A range taken over three levels has more opportunity to be large than a range taken over two, because it is the maximum of three pairwise differences rather than a single one. Any bias in the design therefore inflates the microphysics term relative to the cumulus term, which means the ratio reported here is if anything an underestimate of the true disparity. The ratios quoted throughout should be read as comparative sensitivity ratios for this experiment set — a statement about how much each namelist choice moved this model, on this event, at this resolution — and not as universal physical constants.

The two main effects are close to additive. Fitting the additive model of Sect. 2.7 and taking the residual in each of the six cells gives interaction residuals no larger than 0.29 mm day⁻¹, with a range of 0.57 mm day⁻¹. That is 2.4 times smaller than the microphysics main effect and 18 times smaller than the cumulus main effect. The effect of changing microphysics is therefore much the same whichever cumulus scheme is in use, and the effect of changing cumulus much the same whichever microphysics scheme is in use. This bears directly on how the main effects should be read: because the design is close to additive, they describe separable properties of the two parameterization groups rather than artefacts of the particular six combinations tested, and the ratio between them is not an accident of which pairs happened to be run. With one realisation per cell the interaction cannot be assigned a significance level, but a term this small relative to both main effects gives no support to an interpretation in which the two groups are strongly coupled at this grid spacing.

**Table 4** Domain-mean bias (mm day⁻¹) for the 3 × 2 factorial.

| Microphysics | Kain–Fritsch | Tiedtke | Marginal mean |
|---|---|---|---|
| WSM6 | +9.57 | −0.45 | +4.56 |
| Thompson | +10.99 | −0.10 | +5.45 |
| WDM6 | +9.32 | −1.13 | +4.10 |
| Marginal mean | +9.96 | −0.56 | +4.70 |

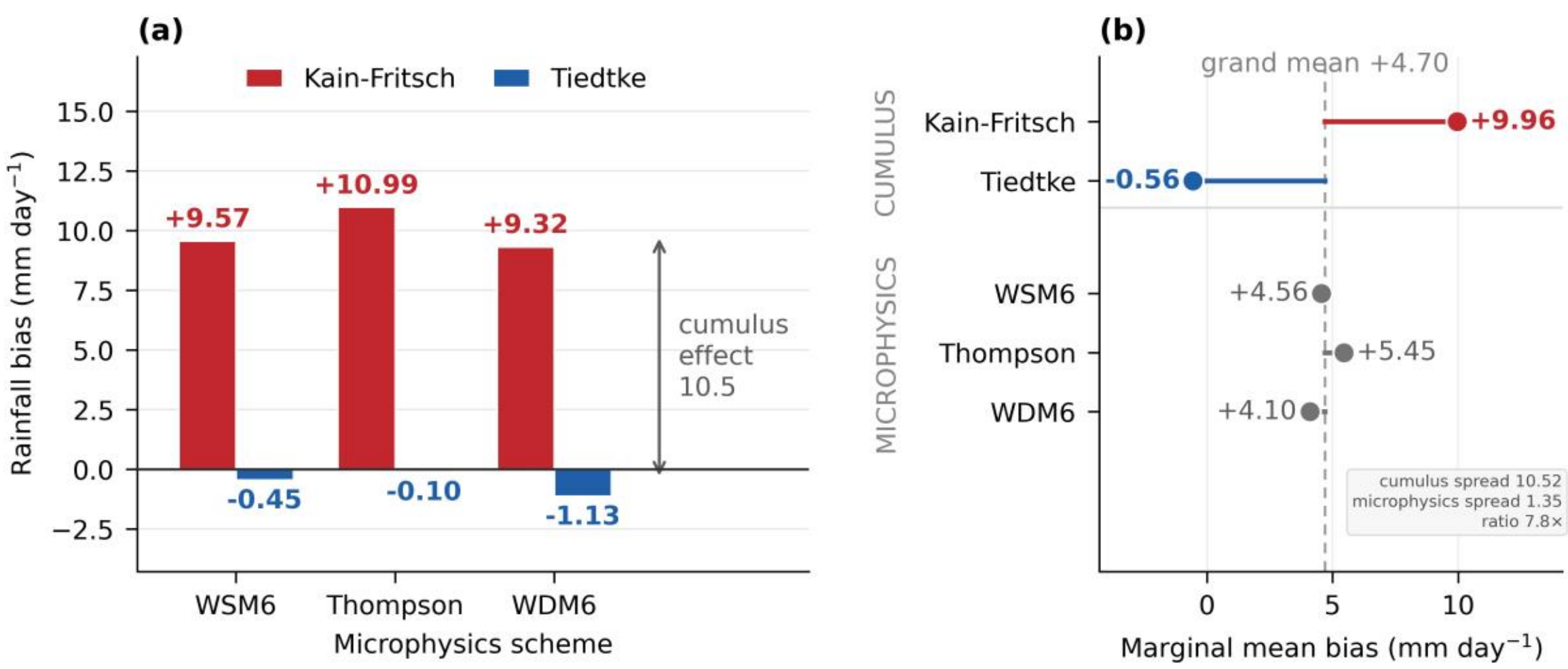

**Fig. 2** Domain-mean rainfall bias for the 3 × 2 factorial. **a** grouped by microphysics scheme and coloured by cumulus scheme; **b** marginal means for each factor about the grand mean

The same ordering holds for daily spatial correlation and for heavy-rainfall detection, but not equally for every score (Table 5). RMSE is the exception and should not be presented as supporting the main result. Recomputed across the processing variants of Sects. 2.4 and 2.5, the cumulus-to-microphysics ratio for RMSE takes values of 0.83, 1.05 and 1.78 depending on the temporal alignment applied, whereas the bias ratio stays within 7.8–8.4 and the correlation ratio within 3.5–4.2. With the adopted +20 h alignment the two main effects on RMSE are of comparable size, 2.01 against 1.91. We therefore conclude that the cumulus scheme dominates bias, daily spatial correlation and heavy-rainfall detection, and that for RMSE the two parameterization groups have effects of similar magnitude.

**Table 5** Main effects for the 3 × 2 factorial.

| Metric | Cumulus main effect | Microphysics main effect | Ratio |
|---|---|---|---|
| Bias (mm day$^{-1}$) | 10.52 | 1.35 | 7.8 |
| Daily spatial r | 0.107 | 0.026 | 4.2 |
| POD (35 mm day$^{-1}$) | 0.155 | 0.032 | 4.8 |
| RMSE (mm day$^{-1}$) | 2.01 | 1.91 | 1.1 |

Kain–Fritsch has much the larger bias but the better daily correlation, 0.441 against 0.334, and substantially better detection of heavy rainfall, with a probability of detection at 35 mm day$^{-1}$ of 0.681 against 0.526. Its mean RMSE is also lower, 35.25 against 37.26 mm day$^{-1}$, though as noted above that margin is not robust to processing choices. Its wet bias is therefore an amplitude error superimposed on a better-located rainfall field, not a symptom of a poorly structured one.

### 3.3 Fractions Skill Score

Base rates and useful-skill targets for the verification period were $f_0$ = 0.790, 0.506, 0.324, 0.198 and 0.147 at the 1, 10, 20, 35 and 50 mm day$^{-1}$ thresholds, giving targets of 0.895, 0.753, 0.662, 0.599 and 0.573 respectively. Fig. 3 shows the Fractions Skill Score against neighbourhood width for all nine experiments.

The smallest neighbourhood at which the score exceeds the useful-skill target is given in Table 6. With correct temporal alignment, all nine configurations attain useful skill at a neighbourhood width of three grid points, approximately 83 km, at 20, 35 and 50 mm day$^{-1}$, and the four Tiedtke-family runs require five points only at 1 and 10 mm day$^{-1}$. The skilful scale is therefore a weak discriminator between schemes in this event once the verification is properly aligned, and it is reported for completeness rather than as a primary result.

**Table 6** Smallest neighbourhood width (grid points) at which the Fractions Skill Score exceeds the useful-skill target.

| Exp | Cumulus | 1 | 10 | 20 | 35 | 50 |
|---|---|---|---|---|---|---|
| 1 | Kain–Fritsch | 3 | 3 | 3 | 3 | 3 |
| 6 | Kain–Fritsch | 3 | 3 | 3 | 3 | 3 |
| 7 | Kain–Fritsch | 3 | 3 | 3 | 3 | 3 |
| 2 | Tiedtke | 5 | 5 | 3 | 3 | 3 |
| 4 | Tiedtke | 5 | 5 | 3 | 3 | 3 |

| Exp | Cumulus | 1 | 10 | 20 | 35 | 50 |
|---|---|---|---|---|---|---|
| 5 | Tiedtke | 5 | 5 | 3 | 3 | 3 |
| 8 | New Tiedtke | 5 | 5 | 3 | 3 | 3 |
| 3 | Grell–Freitas | 3 | 3 | 3 | 3 | 3 |
| 9 | Grell 3D ens. | 3 | 3 | 3 | 3 | 3 |

The score values themselves discriminate more usefully (Table 7). The sign of the cumulus effect reverses with threshold. At the 20 mm day$^{-1}$ threshold Kain–Fritsch is better at every neighbourhood width, with the advantage largest at the grid scale, +0.065, and decaying to +0.015 at nine points. At 35 mm day$^{-1}$ the advantage is negligible at the grid scale, +0.005, and reverses beyond it: Tiedtke is higher at every width of three points or more, by up to 0.033.

This reversal applies to the Fractions Skill Score and not to every heavy-rainfall metric, and the distinction matters because the two can point in opposite directions at the same threshold. At 35 mm day$^{-1}$ the probability of detection continues to favour Kain–Fritsch, 0.681 against 0.526 (Table 5), while the Fractions Skill Score at neighbourhood widths of three points or more favours Tiedtke. The two statements are consistent, because the scores measure different things. The probability of detection asks only whether heavy rainfall was produced where it was observed, and rewards a scheme that produces a great deal of it. The Fractions Skill Score compares the *fraction* of points exceeding the threshold within matched neighbourhoods, so the Kain–Fritsch over-production of heavy rainfall, which the probability of detection rewards, is penalised as the neighbourhood grows and the fractions are averaged over more points. Any statement about the cumulus effect at heavy thresholds must therefore name the score it refers to.

The microphysics spread is included in Table 7 because it changes the interpretation. At 20 mm day$^{-1}$ it is 0.002–0.006, an order of magnitude below the cumulus effect and within the noise of the score. At 35 mm day$^{-1}$ it is 0.011–0.018 and therefore comparable to, or larger than, the cumulus effect. The dominance of the cumulus scheme in this event is thus confined to light and moderate rainfall; at heavy thresholds neither parameterization group dominates and the two are of similar importance.

**Table 7** Cumulus main effect on the Fractions Skill Score, factorial subset, with the microphysics spread for comparison.

| Threshold (mm day$^{-1}$) | Width | Kain–Fritsch | Tiedtke | Difference | Microphysics spread |
|---|---|---|---|---|---|
| 20 | 1 | 0.598 | 0.533 | +0.065 | 0.002 |
| 20 | 3 | 0.723 | 0.682 | +0.041 | 0.006 |
| 20 | 5 | 0.769 | 0.744 | +0.025 | 0.003 |
| 20 | 7 | 0.803 | 0.785 | +0.018 | 0.004 |
| 20 | 9 | 0.828 | 0.814 | +0.015 | 0.004 |
| 35 | 1 | 0.512 | 0.507 | +0.005 | 0.018 |
| 35 | 3 | 0.665 | 0.677 | −0.012 | 0.014 |
| 35 | 5 | 0.720 | 0.746 | −0.026 | 0.011 |
| 35 | 7 | 0.757 | 0.790 | −0.033 | 0.011 |
| 35 | 9 | 0.788 | 0.820 | −0.033 | 0.011 |

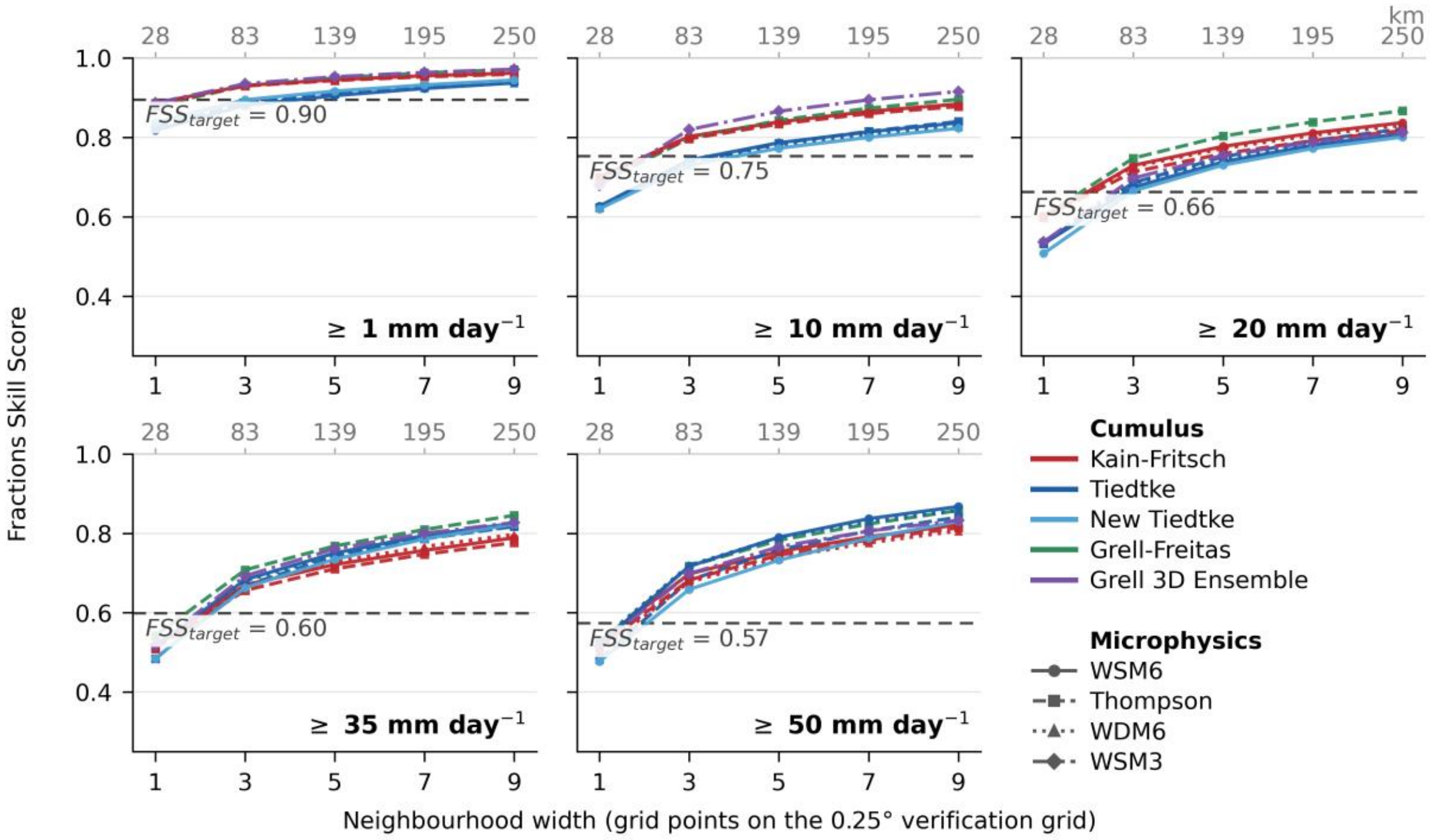

**Fig. 3** Fractions Skill Score against neighbourhood width for all nine experiments, one panel per rainfall threshold. Line colour denotes the cumulus scheme and line style the microphysics scheme. The dashed line marks the useful-skill target, and the upper axis gives the neighbourhood width in kilometres

No cumulus-to-microphysics ratio is quoted for the Fractions Skill Score. Ratios computed against a denominator of 0.002–0.006 are numerically large but not meaningful, and would invite a false impression of precision. The physically interpretable statement is the effect size itself: approximately +0.04 at 20 mm day$^{-1}$ with a 3 × 3 neighbourhood, and approximately −0.01 to −0.03 at 35 mm day$^{-1}$ at widths of three points or more.

### 3.4 The most balanced configuration

Experiment 3, Thompson microphysics with the Grell–Freitas cumulus scheme, showed the most balanced performance among the nine configurations tested. Its bias is +1.94 mm day$^{-1}$, an order of magnitude smaller than any Kain–Fritsch run; its RMSE of 34.26 mm day$^{-1}$ is the third lowest of the nine, behind experiments 9 and 7; its daily correlation of 0.416 is close to the Kain–Fritsch values and well above the Tiedtke values; and it attains the highest Fractions Skill Score of any experiment at both 20 mm day$^{-1}$, 0.748 with a 3 × 3 neighbourhood, and 35 mm day$^{-1}$, 0.708. It therefore achieves Kain–Fritsch-like skill without the Kain–Fritsch wet bias.

It is not the best configuration on every individual score, and no formal combined ranking is defined here. Experiment 9 has the lowest RMSE of the nine, and experiment 6 the highest daily correlation and the highest probability of detection at 35 mm day$^{-1}$. Experiment 3 is singled out because it is the only configuration that is close to the leader on all four score families at once, whereas each of the others is strong on some and weak on others. A different weighting of the scores would select a different configuration, and a user whose application depends primarily on one of them should choose accordingly.

This is a practical rather than a mechanistic result. Grell–Freitas is a scale-aware scheme, which may be why it performs well at a grid spacing in the convective grey zone, but the present experiment set does not isolate scale-awareness as the cause.

### 3.5 The non-factorial experiments

Experiment 9, WSM3 with the Grell 3D ensemble scheme, has the lowest RMSE of the nine at 29.83 mm day$^{-1}$ and a respectable daily correlation of 0.401. This should not be read as the best forecast. It is also the driest run, with a bias of −5.01 mm day$^{-1}$, and it has by far the poorest detection of heavy rainfall, with a probability of detection at 35 mm day$^{-1}$ of 0.412 against 0.669–0.699 for the Kain–Fritsch runs. Its low RMSE follows from producing less rainfall overall rather than from locating rainfall better, a reminder that RMSE alone rewards under-forecasting of a heavy-tailed field.

Experiment 8, New Tiedtke with WSM6, behaves closely like the standard Tiedtke runs on every measure: its bias of −1.25 mm day$^{-1}$ sits just outside the −1.13 to −0.10 mm day$^{-1}$ range of the three Tiedtke runs, and its daily correlation of 0.308 falls within theirs. This is consistent with the two schemes sharing a mass-flux formulation and a common lineage, and it is a useful negative result: a change of scheme version within the same family moves the solution far less than a change of family.

Experiment 3, Thompson with Grell–Freitas, is the one non-factorial run that does not resemble either factorial group. Its bias of +1.94 mm day$^{-1}$ lies between the two, but its daily correlation of 0.416 and its detection of heavy rainfall at 0.559 are closer to the Kain–Fritsch values than to the Tiedtke ones, and its Fractions Skill Score is the highest of any experiment at both 20 and 35 mm day$^{-1}$. It is discussed further in Sect. 3.4.

Taken together the three non-factorial runs serve two purposes. They test whether the factorial result is an artefact of the particular pair of cumulus schemes crossed: it is not, since the New Tiedtke run reproduces the Tiedtke behaviour and both Grell-family runs depart from both groups, so the spread across cumulus schemes is if anything wider than the factorial alone indicates. They also show that the cumulus effect is not one-dimensional. The five cumulus schemes do not lie on a single wet-to-dry axis: experiment 9 is dry and poor at heavy rainfall, experiment 3 is near-neutral and good at it, and the two cannot be ordered by bias alone.

### 3.6 Spatial structure

Fig. 4 shows the difference between the simulated and observed event totals for 7–12 July 2023, arranged so that rows are cumulus schemes and columns microphysics schemes; the totals themselves are given in the Supplementary Information (Fig. S1). The figure has eight panels: the observed field, the six factorial experiments, and the non-factorial experiment 9, giving seven model panels in all. Within each row the three microphysics columns are visually almost indistinguishable, in Fig. 4 and in Fig. S1 alike, which is the factorial result of Sect. 3.2 in spatial form.

The difference fields separate into two distinct signals. The first is a northern over-prediction common to every configuration: a coherent band of positive difference lies along approximately 32–33.5°N, 75.5–78°E, reaching +300 mm in event total. It is present in all seven model panels, including the three Tiedtke runs and experiment 9, which are close to neutral or dry in the domain mean. This signal is therefore not attributable to the cumulus scheme, because every configuration produces it.

Experiments 3 and 8 are not shown in Fig. 4, whose layout displays the 3 × 2 factorial together with one non-factorial run in the free corner slot. Their domain-mean biases, +1.94 and −1.25 mm day$^{-1}$ respectively (Table 3), place them between the Kain–Fritsch and Tiedtke groups, and their difference fields show the same northern over-prediction as every other configuration.

The second is a central deficit confined to the Tiedtke-family runs. Over roughly 29–31.5°N, 76.5–79.5°E, which is where the observed event maximum lies, the Tiedtke runs and experiment 9 are strongly dry, locally by more than 300 mm, whereas the Kain–Fritsch runs are close to neutral or slightly wet in the same region. The cumulus signal is thus spatially localised rather than uniform, and it is largest precisely where the observed rainfall is heaviest.

This resolves an apparent tension in the domain-mean statistics. The Tiedtke runs achieve near-zero domain-mean bias not by reproducing the observed field but by compensating a large northern excess against a large central deficit. The Kain–Fritsch runs carry the same northern excess but additionally fill the observed maximum, which produces a positive domain-mean bias while also producing the higher daily correlation and the substantially better heavy-rainfall detection reported in Sect. 3.2. A near-zero area-average bias is therefore a poor summary of skill for this event, and the spatial fields are needed to interpret Table 3 correctly.

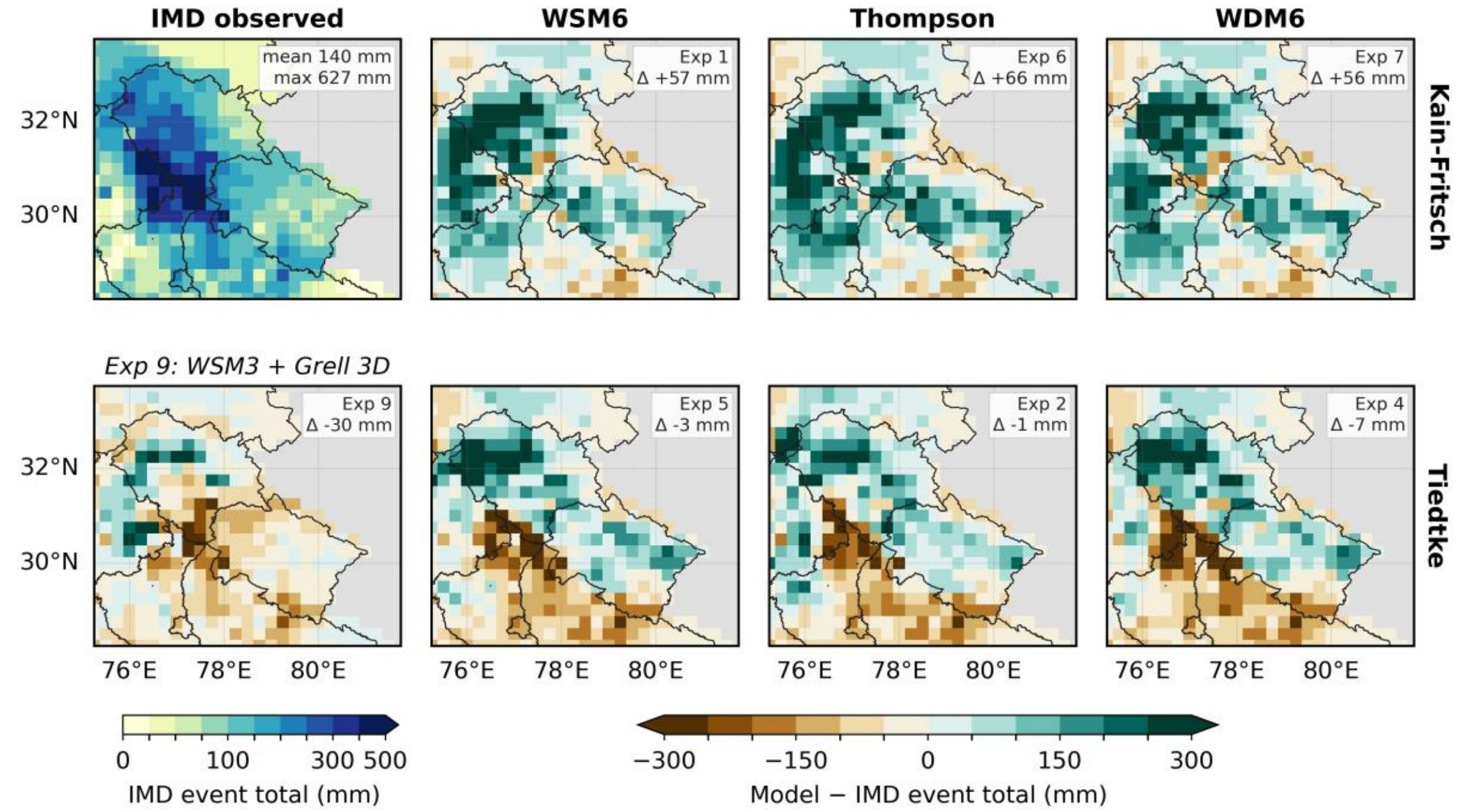

**Fig. 4** Simulated minus observed event total for 7–12 July 2023. Rows denote the cumulus scheme and columns the microphysics scheme, with the observed field and the non-factorial experiment 9 in the left column. The observed field uses the left colour bar and the differences the right one

The common northern signal warrants caution rather than interpretation as pure model error. Fig. 5 shows that on 8–9 July IMERG also places substantial rainfall in this part of the domain, as do ERA5 and the GFS analysis, not shown. The region lies at high elevation where the IMD gauge network is sparse, so an unknown fraction of the apparent over-prediction may reflect under-observation rather than model error. No attempt is made to partition it here; verification against an independent satellite reference would be required, and is identified as the natural next step in Sect. 4.4.

Fig. 5 also provides a direct check on the temporal alignment of Sect. 2.4. With the +20 h shift applied, IMD and all three plotted experiments attain their domain-mean maximum on the same day, 9 July. IMERG peaks one day earlier, on 8 July, because it is accumulated on the UTC day and no shift has been applied to it; this is the expected behaviour given the two conventions.

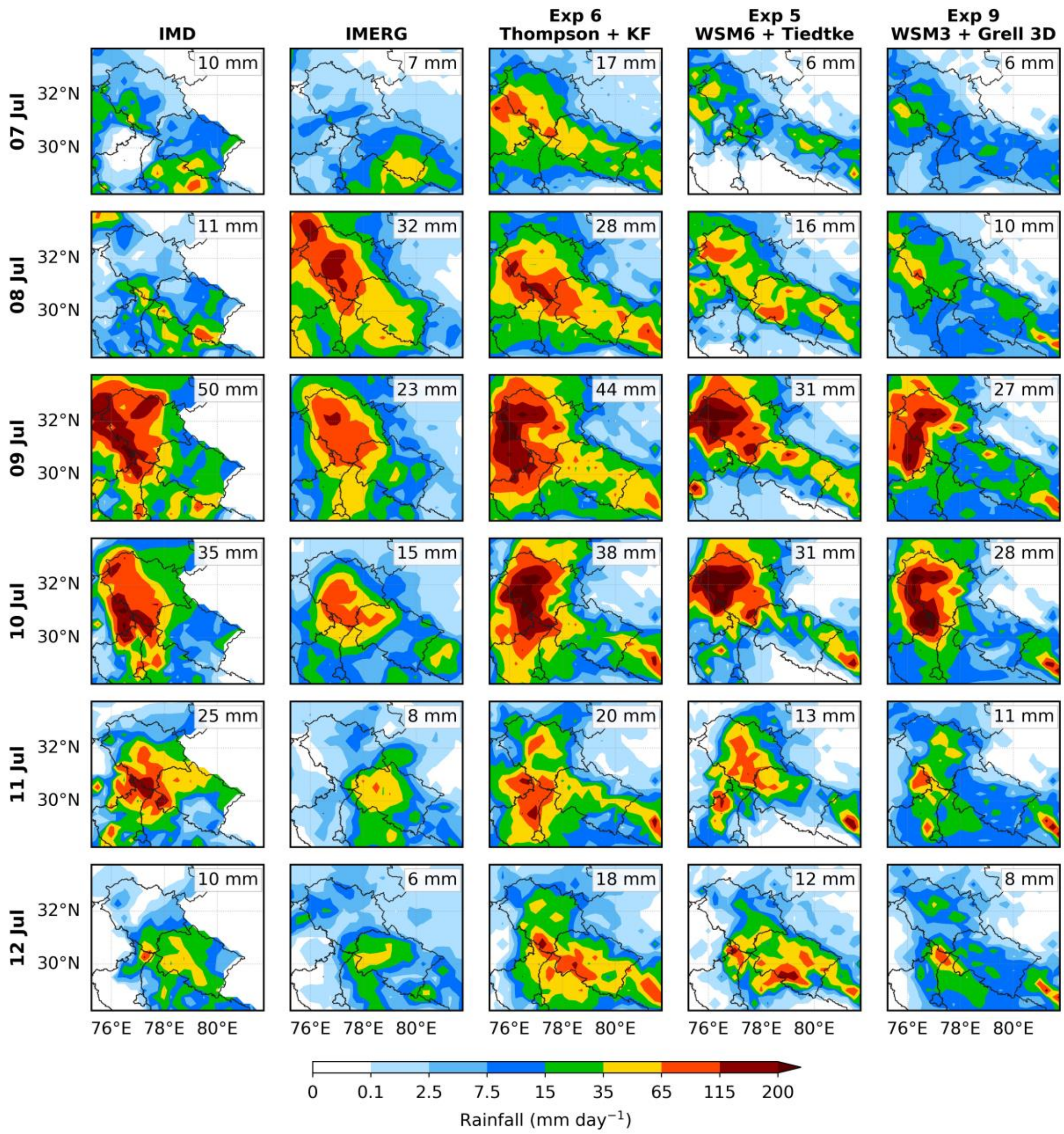

**Fig. 5** Daily rainfall for 7–12 July 2023 from IMD, IMERG and three representative experiments: experiment 6 (Thompson with Kain–Fritsch), experiment 5 (WSM6 with Tiedtke) and experiment 9 (WSM3 with the Grell 3D ensemble scheme). IMD and the model fields are accumulated on the IMD reporting day, 0300–0300 UTC; IMERG is accumulated on the UTC calendar day and therefore appears one day earlier

## 4. Discussion

The results raise three questions, which this section takes in turn. Section 4.1 asks why the cumulus scheme should dominate so heavily at this grid spacing, and argues that the reason is structural rather than a matter of one physical process happening to be stronger than another. Section 4.2 sets out what that implies for anyone configuring a regional model over this terrain, and reconciles the present result with recent microphysics-sensitivity studies over Indian domains, including one addressing this same event. Section 4.3 returns to the verification methodology of Sects. 2.4 and 2.5, whose effect on the scores proved comparable to the physics effect being measured. Section 4.4 states the limitations of the experiment set.

### 4.1 Why the cumulus scheme dominates

The disparity between the two main effects has a structural explanation as well as a physical one, and the structural explanation is the more complete of the two.

In WRF the total surface rainfall is the sum of two separately computed components. The convective component is produced within the cumulus scheme, which determines when convection is triggered, how much mass is transported, how much of the condensate it generates is precipitated and how much is detrained into the grid-scale fields. That precipitation falls out inside the scheme and is accumulated in its own output field. It does not pass through the microphysics scheme at any stage. The grid-scale component is the one the microphysics scheme computes, acting on condensate produced by the resolved dynamics or detrained into the grid-scale fields by the cumulus scheme.

The consequence is not a matter of degree. To the extent that the convective component carries the total, the microphysics scheme has no pathway through which to influence it. Changing from WSM6 to Thompson to WDM6 alters the treatment of detrained condensate and of whatever the resolved ascent produces, but leaves the convectively generated rainfall untouched. This is why the microphysics main effect is small rather than merely smaller: it is confined to a fraction of the total by the structure of the model, not by the relative strength of two competing physical processes. The near-additivity reported in Sect. 3.2 fits the same picture. If the two groups were competing for control of the same rainfall, their effects would be expected to interfere; an interaction term 18 times smaller than the cumulus main effect is what a largely one-way dependence looks like.

The threshold behaviour follows from the same argument and provides a test of it. If cumulus dominance arises because the convective component carries most of the rainfall, then the dominance should weaken wherever the resolved component carries more of it. Heavy rainfall is where that is expected: a cumulus scheme parameterizes sub-grid convection and returns a smoothed field, whereas the largest accumulations at 9 km are more likely to arise from resolved ascent over terrain, which the microphysics scheme controls directly. This is what Table 7 shows. At 20 mm day$^{-1}$ the microphysics spread in the Fractions Skill Score is 0.002–0.006, negligible against a cumulus effect of 0.015–0.065; at 35 mm day$^{-1}$ it rises to 0.011–0.018 while the cumulus effect falls to between +0.005 and −0.033, so that the two become comparable and the cumulus effect changes sign. The balance between the two groups therefore shifts systematically with rainfall intensity, in the direction the structural argument predicts.

Within the convective component, the difference between Kain–Fritsch and Tiedtke is one of trigger and closure rather than of formulation type, since both are mass-flux schemes. The spatial fields of Sect. 3.6 indicate where the two diverge. They agree over the northern part of the domain, where both produce a substantial excess, and differ over the central belt at approximately 29–31.5°N, 76.5–79.5°E, where the observed event maximum lies and where the Tiedtke runs are dry by more than 300 mm in event total while the Kain–Fritsch runs are close to neutral. The difference between the two schemes is therefore not a uniform scaling of convective activity across the domain but is concentrated where the environment is most favourable to deep convection, which is consistent with the schemes responding differently to strong instability rather than differing in their background behaviour.

This account is inferred from the total rainfall fields rather than demonstrated directly, and the distinction should be stated plainly. A direct test is available: WRF accumulates the convective and grid-scale components separately, so partitioning the simulated rainfall between them would give the convective fraction explicitly, and would establish whether that fraction falls at heavy thresholds as the argument requires. The partition was not performed here because the analysis was built throughout on total rainfall, for comparability with the IMD product, which makes no such distinction. It is the most direct extension of this work and is recommended for any study that seeks to confirm the mechanism rather than infer it.

### 4.2 Implications for configuration choice

For a practitioner configuring a regional model over this terrain at comparable resolution, the practical implication is direct: effort spent selecting and evaluating the cumulus scheme will yield substantially more improvement than the same effort spent on microphysics. The corollary is that a study reporting microphysics sensitivity while holding one cumulus scheme fixed may be reporting a second-order effect. Several recent studies over Indian terrain take that form (Sharma et al. 2024; Saini et al. 2026), and the present results suggest that their conclusions hold within a cumulus configuration rather than across configurations. This is not a criticism of those studies, which address a different question; it is a statement about what a microphysics-only comparison can establish at this grid spacing.

Saini et al. (2026) examined this same July 2023 event in a convection-permitting framework, with cumulus parameterization switched off, and reported clear sensitivity to the microphysics scheme. That result and the present one are consistent rather than contradictory: removing the cumulus scheme necessarily promotes the microphysics scheme to the dominant control. Read together, the two bracket the problem, and the comparison suggests that the decision between running with a cumulus scheme at grey-zone resolution and running convection-permitting is itself a larger one than any within-group scheme choice.

The threshold reversal qualifies this. A configuration chosen for the best area-average or moderate-rainfall performance is not necessarily the best configuration for extreme rainfall, and for an application concerned with flood-producing rainfall the ranking may invert.

### 4.3 Verification methodology as a source of apparent error

The temporal alignment correction of Sect. 2.4 changed the daily spatial correlation of experiment 1 from 0.375 to 0.437, RMSE from 43.64 to 36.34 mm day$^{-1}$ and the probability of detection at 35 mm day$^{-1}$ from 0.576 to 0.669. The magnitude of that change is comparable to, and for RMSE larger than, the difference between the cumulus schemes on the same metrics.

Any study that compares model rainfall against the IMD gridded product without matching the 0830 IST reporting window is therefore reporting skill scores that are depressed by a methodological artefact, and the depression is not uniform across metrics: bias is barely affected while pattern-based scores are affected substantially. A practical consequence is that comparisons of RMSE between schemes are unreliable unless the alignment is handled, since the cumulus-to-microphysics ratio for RMSE varies from 0.83 to 1.78 across alignment choices in this dataset.

The regridding result carries a similar message. Bicubic interpolation of a strictly non-negative field across sharp orographic gradients produced values as low as −3.9 mm day$^{-1}$, with implied compensating positive errors of similar magnitude in the same region.

### 4.4 Limitations

Several limitations should be stated plainly.

Single event. These are nine simulations of one event. The magnitude of the cumulus-to-microphysics ratio should not be assumed to transfer to other events, seasons or regions without testing.

Domain size. The 720 km domain gives limited room for the model to develop its own mesoscale circulation before the solution is influenced by the lateral boundaries, and with 6-hourly boundary forcing the solution is substantially boundary-controlled. This affects the absolute skill of all runs, but because the domain is identical across experiments it does not affect the comparison between them.

Observational reference. The IMD gridded product is interpolated from a gauge network that is sparse at high elevation, and Sect. 3.6 shows a region where it disagrees with three independent products and with every simulation. Part of the diagnosed wet bias may reflect under-observation rather than model error. Repeating the verification against IMERG as an independent reference would help bound this and is a natural next step.

Gravity wave drag. The gravity wave drag option was used at 9 km, a grid spacing at which the scheme's assumptions are marginal and where some double-counting with resolved orographic effects is possible. The setting is identical across all nine runs and therefore does not affect the intercomparison.

Pooled Fractions Skill Score. The score reported here is aggregated over all verification days rather than computed per day and averaged. This is a standard choice but means a single day with unusual behaviour can influence the aggregate.

## 5. Conclusions

Nine WRF simulations of the 6–13 July 2023 Western Himalayan rainfall event, differing in exactly two namelist options, were verified against IMD gridded rainfall over 447 land points for 7–12 July 2023.

1. For the 9 km WRF configuration and the July 2023 Western Himalayan event examined here, the cumulus parameterization dominates the microphysics parameterization in controlling simulated rainfall. In a controlled 3 × 2 factorial, switching cumulus scheme changed the domain-mean bias 7.8 times more than switching microphysics scheme, with comparative sensitivity ratios of 4.2 for daily spatial correlation and 4.8 for detection of heavy rainfall. These ratios characterise this experiment set rather than a universal property of the two parameterization groups, and a single event cannot establish generality. RMSE is the exception: the two effects are of comparable size and their ratio is not robust to processing choices, so cumulus dominance is not claimed for that score.
2. Kain–Fritsch produces a systematic wet bias but a better-structured rainfall field. Its bias of +9.3 to +11.0 mm day$^{-1}$ is accompanied by higher daily correlation, 0.441 against 0.334, and substantially better heavy-rainfall detection, 0.681 against 0.526, than the near-neutral Tiedtke runs.
3. The cumulus advantage reverses with threshold in the Fractions Skill Score, and the microphysics effect grows. Kain–Fritsch is superior from 1 to 20 mm day$^{-1}$, but at 35 mm day$^{-1}$ Tiedtke attains higher Fractions Skill Scores at every neighbourhood width of three grid points or more, and the microphysics spread becomes comparable to or larger than the cumulus effect. The reversal is specific to that score: at the same threshold the probability of detection still favours Kain–Fritsch, 0.681 against 0.526, because it rewards the production of heavy rainfall where the Fractions Skill Score penalises its over-production. Conclusions about the cumulus effect at heavy thresholds are therefore only meaningful when the score is named.
4. Thompson microphysics with the Grell–Freitas cumulus scheme showed the most balanced performance among the nine configurations tested, combining near-zero bias, +1.94 mm day$^{-1}$, with the highest Fractions Skill Score of any experiment at both moderate and heavy thresholds. It does not lead on every score — experiment 9 has the lowest RMSE and experiment 6 the highest correlation and probability of detection — and no formal combined ranking is defined here.

5. Verification methodology accounts for a portion of apparent model error comparable to the difference between physics schemes. Matching the model accumulation window to the IMD 0830 IST reporting day raised daily spatial correlation from 0.375 to 0.437, lowered RMSE from 43.64 to 36.34 mm day$^{-1}$ and raised the probability of detection at 35 mm day$^{-1}$ from 0.576 to 0.669. The optimum occurred at exactly the shift predicted from the documented reporting window and degraded on either side of it. Separately, bicubic regridding produced spurious negative rainfall of up to −3.9 mm day$^{-1}$ that bilinear regridding eliminates.

## Statements and Declarations

**Funding** No funding was received for conducting this study.

**Competing interests** The authors have no relevant financial or non-financial interests to disclose.

**Data availability** The IMD 0.25° × 0.25° gridded daily rainfall dataset is publicly available from the India Meteorological Department, Pune (Pai et al. 2014; https://www.imdpune.gov.in/cmpg/Griddata/Rainfall_25_NetCDF.html). The GPM IMERG Final Run daily product is publicly available from the NASA Goddard Earth Sciences Data and Information Services Center (Huffman et al. 2023). The NCEP FNL operational global analyses used for initial and lateral boundary conditions are available from the NSF NCAR Research Data Archive (NCEP 2000). **Code availability** All material generated for this study has been deposited in a public repository: the nine WRF namelists, the pre-processing and temporal-alignment scripts, the regridding commands, the derived daily rainfall fields of every experiment on the 0.25° verification grid, and the code that computes the verification statistics, the factorial main effects and the Fractions Skill Score, together with the plotting scripts for every figure. The repository address is withheld from this version of the manuscript because it identifies the authors under the journal's double-anonymous policy. It will be supplied to the editor on request and stated in full in the published version.